# Modeling the Seasonal Variability of the Plasma Environment in Saturn's Magnetosphere between Main Rings and Mimas

**W.-L. Tseng, R. E. Johnson and M. K. Elrod**

**University of Virginia, Charlottesville, VA 22904, USA**

*Corresponding author: W.-L. Tseng wt7b@virginia.edu*

**Abstract:**

The detection of $O_2^+$ and $O^+$ ions over Saturn's main rings by the Cassini INMS and CAPS instruments at Saturn orbit insertion (SOI) in 2004 confirmed the existence of the ring atmosphere and ionosphere. The source mechanism was suggested to be primarily photolytic decomposition of water ice producing neutral $O_2$ and $H_2$ (Johnson et al., 2006). Therefore, we predicted that there would be seasonal variations in the ring atmosphere and ionosphere due to the orientation of the ring plane to the sun (Tseng et al., 2010). The atoms and molecules scattered out of the ring atmosphere by ion-molecule collisions are an important source for the inner magnetosphere (Johnson et al., 2006; Martens et al. 2008; Tseng et al., 2010 and 2011). This source competes with water products from the Enceladus' plumes, which, although possibly variable, do not appear to have a seasonal variability (Smith et al., 2010). Recently, we found that the plasma density, composition and temperature in the region from 2.5 to 3.5 $R_S$ exhibited significant seasonal variation between 2004 and 2010 (Elrod et al., 2011). Here we present a one-box ion chemistry model to explain the complex and highly variable plasma environment observed by the CAPS instrument on Cassini. We combine the water products from Enceladus with the molecules scattered from a corrected ring atmosphere, in order to describe the temporal changes in ion densities, composition and temperature detected by CAPS. We found that the observed temporal variations are primarily seasonal, due to the predicted seasonal variation in the ring atmosphere, and are consistent with a compressed magnetosphere at SOI.

## 1. Introduction

Saturn's oxygen ring atmosphere and ionosphere was discovered by the Cassini spacecraft in 2004. Both the Ion and Neutral Mass Spectrometer (INMS) and the Cassini Plasma Analyzer (CAPS) detected $O_2^+$ and $O^+$ ions over the main rings during the Saturn Orbital Insertion (SOI) in 2004 (Tokar et al., 2005; Waite et al., 2005) as well as a comparable component of electrons (Coates et al., 2005). Johnson et al. (2006) suggested that the primary source is photolytic decomposition of water ice producing neutral $O_2$ and $H_2$. The measurements also indicated that the ion-molecule collisions between the newly formed ions $O_2^+/O^+$ and neutral $O_2$ determined the ion density distribution detected by CAPS at high altitudes above the main rings. Using the photolytic model we also predicted that there would be significant seasonal variations in the density of the ring atmosphere and ionosphere since the neutral $O_2$ production rate depends on the solar incident angle with respect to the ring plane (Tseng et al., 2010). The ion-molecule collisions, which account for the observed vertical distribution, also scatter $O_2$, $H_2$, O and H beyond the ring plane into the magnetosphere and into Saturn's atmosphere. Once ionized, they are a source of $O_2^+$ and $H_2^+$ ions in the magnetosphere seen by Cassini Magnetospheric Imaging Instrument (MIMI) (Krimigis et al., 2005) and CAPS (Martens et al., 2008; Tseng et al., 2011). Therefore, the seasonal variations should also be reflected in the density of these ions in the region outside the main rings.

However, the suggested seasonal variations are complicated by the deposition of water products from the Enceladus' plumes onto the A-ring as described by measurements and modeling (Jurac and Richardson, 2007; Farrell et al., 2008; Cassidy and Johnson, 2010). As with the dissociated oxygen in the ring atmosphere, the absorbed oxygen rich ions and neutrals from the Enceladus torus could be recycled via grain-surface chemistry contributing to the atmosphere over the main rings and the atoms and molecules scattered into the magnetosphere (Tseng and Ip, 2011). Hansen et al. (2011) monitored the Enceladus plume activity using the Cassini Ultraviolet Imaging Spectrograph (UVIS) and found that they appeared to be stable. Although the INMS measurements suggested the source rate is variable (Smith et al., 2010), no seasonal variability was apparent. So, the Enceladus torus contribution to the ring atmosphere can mitigate the seasonal variation in the ring atmosphere. Despite the uncertainties in the recycling efficiency, the non-detection (or upper limit) of $H_2^+$ ions over the B-ring by the Cassini CAPS at SOI has helped constrain the source rates (Tseng et al., 2011).

Elrod et al. (2011) examined the CAPS plasma data between 2.5 and 3.5 $R_S$ from 2004 to 2010 including a comparison with Voyager 2 data from Richardson (1986) where $R_S$ is one mean Saturn radius (60,300 km). They showed that there were large variations over that time period in the ion density, temperature and composition. A significant drop in the ion density and temperature was found between 2005 and 2010 as compared to Voyager 2 data and Cassini data at SOI (2004). They also found that the $O_2^+$ was the dominant heavy ion at SOI, but the $O_2^+$ density was comparable to or less than the water-group ion (hereafter referred to as $W^+$) density in the period of 2005-2010. Using preliminary results from the model described in detail here, they concluded that, although the possible variability in the Enceladus source might contribute, the observed variations were primarily seasonal due to the predicted seasonal variation in the ring atmosphere.

In this work, we developed a one-box ion chemistry model for the plasma in the region between 2.5 and 3.5 $R_S$ to explain the complex and highly variable plasma environment that was detected by CAPS. First, an updated ring atmosphere model is described in Section 2, since this atmosphere is affected by the recently observed change in the ring particle temperatures from solstice to equinox which was absent in our earlier model (Tseng et al., 2010). In Section 3, our ion-chemistry model, combining the Enceladus torus and the seasonal ring atmosphere sources, is presented. The simulations are carried out for both SOI and equinox conditions and the results are in remarkable agreement with the trends found in the CAPS data. In Section 4, we solved a set of continuity equations for the plasma energy in order to understand the low ion temperature detected after SOI. It is found that ion energy lost to momentum transfer during ion-molecule collisions can roughly account for decrease of ion temperature. Finally, a summary is given in Section 5.

## 2. Revised ring atmosphere model: neutral source rates of $O_2$ and $H_2$ at SOI and equinox

In our previous paper (Tseng et al., 2010), we studied the structure and time variability of Saturn's ring atmosphere and ionosphere using a fixed ring temperature T=100K accounting only for the change in solar UV flux on the $O_2$ production rate. However, the average temperature of Saturn's ring particles also changes significantly with the change in the solar incident angle as revealed by the Cassini Composite Infrared Spectrometer (CIRS) data (Flandes et al., 2010): T~100 K at solstice and T ~ 60 K at equinox. Since this change primarily occurs for the A- and B-ring particles, which are the principal source of the ring atmosphere, the photo-production of $O_2$ and $H_2$ and recycling on the ring particles are significantly modified by this change in the particle temperatures. Therefore, we re-examine the time variability of Saturn's ring atmosphere and ionosphere allowing for the effect of solar illumination angle and the ring particle temperature. In addition we consider the influence of the deposition of oxygen from Enceladus torus onto the A-ring.

$O_2$ production by UV photons depends on the UV flux and, hence, on the solar incident angle with respect to the ring plane as well as the ring particle temperature. In our earlier work in 2010, we used $Q(O_2) = c \times 10^6$ molecules $cm^{-2}$ $s^{-1}$ with a fixed ring temperature T= 100 K. This was based on the model in Johnson et al. (2006) to describe the observed plasma densities primarily over the B-ring with the parameter c accounting for recycling of dissociated oxygen and oxygen ions on the ring particle surfaces. At T=100K and solar incident angle $\gamma = 24$ degrees the neutral $O_2$ source rate was computed from laboratory data to be ~ $1.0 \times 10^{26}$ $s^{-1}$ at SOI due to photolytic decomposition of water ice. Since the calculated ion densities were roughly a factor of 20 smaller than our recent re-analysis of the CAPS data over the B-ring at SOI (Elrod et al., 2011), the effective source at SOI was estimated to be ~$2.0 \times 10^{27}$ $s^{-1}$ (c ~ 20) accounting for recycling (e.g. Johnson et al., 2006), the influences of Enceladus' plumes (Tseng and Ip, 2011), the solar activity and, possibly, the simplified ion-molecule interactions.

Earlier we estimated a photolytic $O_2$ source rate ~$1.0 \times 10^{25}$ $s^{-1}$ near equinox again assuming an average ring particle temperature of T=100 K (Tseng et al., 2010). Using the same model but with an average temperature of ~60 K near equinox (Flandes et al., 2010), the photolytic source rate is severely quenched by a factor of $\exp(-\alpha/kT)$ where $\alpha$ is an activation barrier, which is somewhat uncertain. Here we use $\alpha \sim 0.03$ eV for the photo-production of $O_2$ (e.g., Johnson, 2011). Using T=60K as the average temperature of the particles gives a photolytic yield a factor of ~10 reduction, in addition to the decrease due to the solar illumination angle, or ~ $1.0 \times 10^{24}$ $s^{-1}$. In absence of CAPS data over the main rings near equinox, we allow that recycling of desorbed oxygen and the contribution of oxygen from Enceladus could also enhance this source rate by the same factor as summarized in Table 1. Fortunately, the conclusions discussed below are not critically dependent on the exact size of the equinox ring atmosphere source rate, but only on the fact that the source rate is significantly smaller than that at SOI.

The momentum transfer during collisions between primarily $O_2^+$ ions and neutral $O_2$ molecules creates a component of $O_2$ molecules scattered into the magnetosphere (Johnson et al., 2006; Martens et al., 2008; Tseng et al., 2010). Figure 1 shows the calculated radial profiles of the neutral $O_2$ column density from the rings to a radial distance of 10 $R_S$ for the SOI and equinox source rates and for an intermediate source rate. Once ionized, they contribute to the local magnetospheric plasma. It is seen that the neutral $O_2$ column density over the main rings can be roughly scaled to the $O_2$ source rate. But, outside the main rings, the scattered population formed by ion-molecule collisions changes nearly quadratically with the source rate, as it depends on both the neutral and the ion density. That is, when the $O_2$ source rate becomes ten percent of SOI source rate, the scattered $O_2$ column density outside the main rings drops to be roughly one percent of SOI value.

The azimuthally symmetric spatial distributions of the $O_2^+$ ions produced by photoionization over the main rings are shown in Figure 2 for (a) equinox and (b) SOI phase. It is clearly seen that the $O_2^+$ source rate depends on the orientation of the ring plane to the Sun with the $O_2^+$ density significantly lower at equinox. The structure and main features (such as the asymmetry above and below the ring plane) of the ring ionosphere have been explored in Johnson et al. (2006) and Tseng et al. (2010). Although the calculated ion densities involve considerable uncertainties such as the recycling of oxygen on grain surfaces, we note that, if the ring source rate is not overwhelmed by material from Enceladus, the seasonal excursion in the size of Saturn's $O_2$ ring atmosphere and ionosphere combining the effects of solar incident angle and ring temperature might be much larger than predicted earlier (Tseng et al., 2010), with the densities at solstice around two orders of magnitudes higher than near equinox.

Both neutral $O_2$ and $H_2$ molecules are released in a ratio of 1:2 from main rings by decomposition of ice by solar UV photons [Johnson et al., 2006]. We carried out the simulations of the spatial morphology of the ring $H_2$ atmosphere and the ring $H_2^+$ ionosphere including the ion-molecule collisions between $H_2$ ($H_2^+$) and $O_2^+$/ $W^+$ ($O_2$/ W) (Tseng et al., 2011). Figure 3 shows the radial distributions of the neutral $O_2$ and $H_2$ column density with a source rate of ~ $2x10^{26}$ $H_2$ $s^{-1}$ and of ~$1x10^{26}$ $O_2$ $s^{-1}$, respectively. Overall, the neutral $H_2$ column density over the main rings is approximately one order of magnitude larger than $O_2$ since $H_2$ has a longer photolytic lifetime. Compared to $O_2$, $H_2$ attains a larger thermal velocity on thermal desorption from the ring particles and obtains more energy during ion-molecule collisions because of its smaller mass. Therefore, in the region between the main rings and Enceladus, our model of the *column density* of scattered $H_2$ is roughly two orders of magnitudes higher than our model $O_2$ column density but the number density of $H_2$ near the equatorial plane is only roughly one order of magnitude higher. In addition, Tseng et al.

(2011) found that, during periods where the ring source rate of $2x10^{26}$ $H_2$ $s^{-1}$ applies, $H_2$ scattered from the ring atmosphere and photodissociation of water from Enceladus are comparable sources in the magnetosphere between ~6 and ~2.5 $R_S$. At any time, Titan is the dominant source of $H_2$ in the outer magnetosphere. The density distribution of $H_2^+$ outside the main rings estimated from our model (Tseng et al., 2011) roughly agrees with the CAPS observations (Thomsen et al., 2010). In addition, we note that, CAPS should have detected $H_2^+$ in Cassini division if $H_2$ and $O_2$ are formed stoichiometrically (Tseng et al., 2011). Since CAPS did not detect $H_2^+$ over the main rings at SOI (Young et al., 2005; Tokar et al., 2005), it is likely that the same enhancement factor used for $O_2$ recycling does not apply since the H produced by dissociation has a higher velocity and is lost from the region (Johnson et al., 2006). Either non-detection (an upper limit) over the rings or measurement of $H_2^+$ in the magnetosphere can be used to constrain the $H_2/O_2$ source mechanism from main rings.

Table 1: Summary of the neutral $O_2$ and $H_2$ source rate of main rings

| **Source Rate** | **SOI**<br>**Solar incident angle = 24**<br>**Ring T = 100K** | **Equinox**<br>**Solar incident angle = 2**<br>**Ring T = 60K** |
|---|---|---|
| $O_2$ (1) | $1.0x10^{26}$ $s^{-1}$ | $1.0x10^{24}$ $s^{-1}$ |
| $O_2$ (2) | $2.0x10^{27}$ $s^{-1}$ (3) | $2.0x10^{25}$ $s^{-1}$ |
| $H_2$ (1) | $2.0x10^{26}$ $s^{-1}$(4) | $2.0x10^{24}$ $s^{-1}$ |

(1) Photolytic source rate
(2) Corrected for recycling which includes contributions of oxygen from Enceladus deposited on the A-ring (see discussion in Section 2)
(3) Correction based on calibration to the CAPS SOI data (Elrod et al., 2011)
(4) No detection (or upper-limit) of $H_2^+$ to help us constrain the neutral $H_2$ source rate

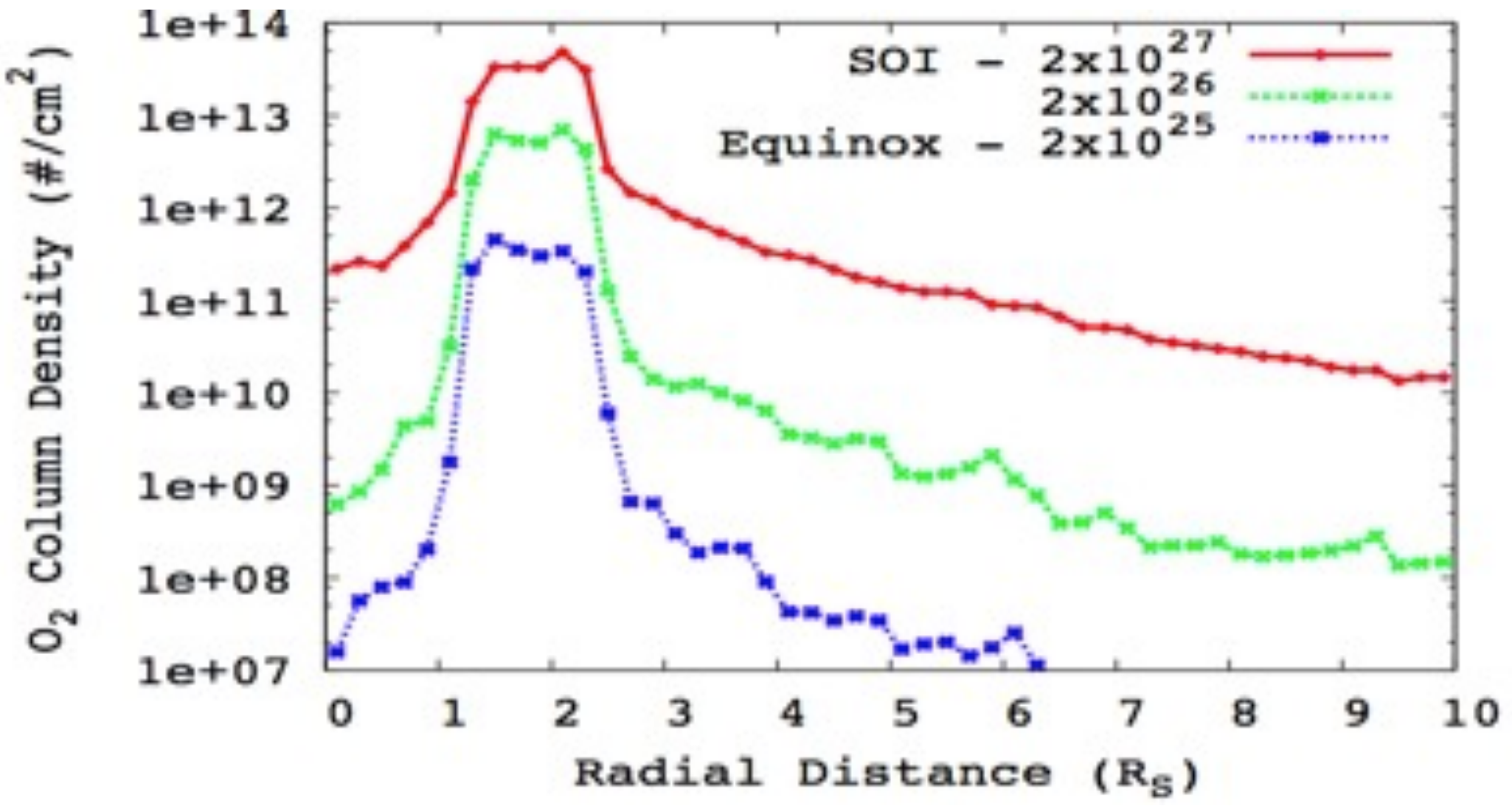

Figure 1: The radial distributions of the neutral $O_2$ column density with a source rate ~ $2x10^{27}$ $O_2$ $s^{-1}$ (SOI phase), ~$2x10^{26}$ $O_2$ $s^{-1}$ (intermediate phase) and ~$2x10^{25}$ $O_2$ $s^{-1}$ (equinox phase), respectively.

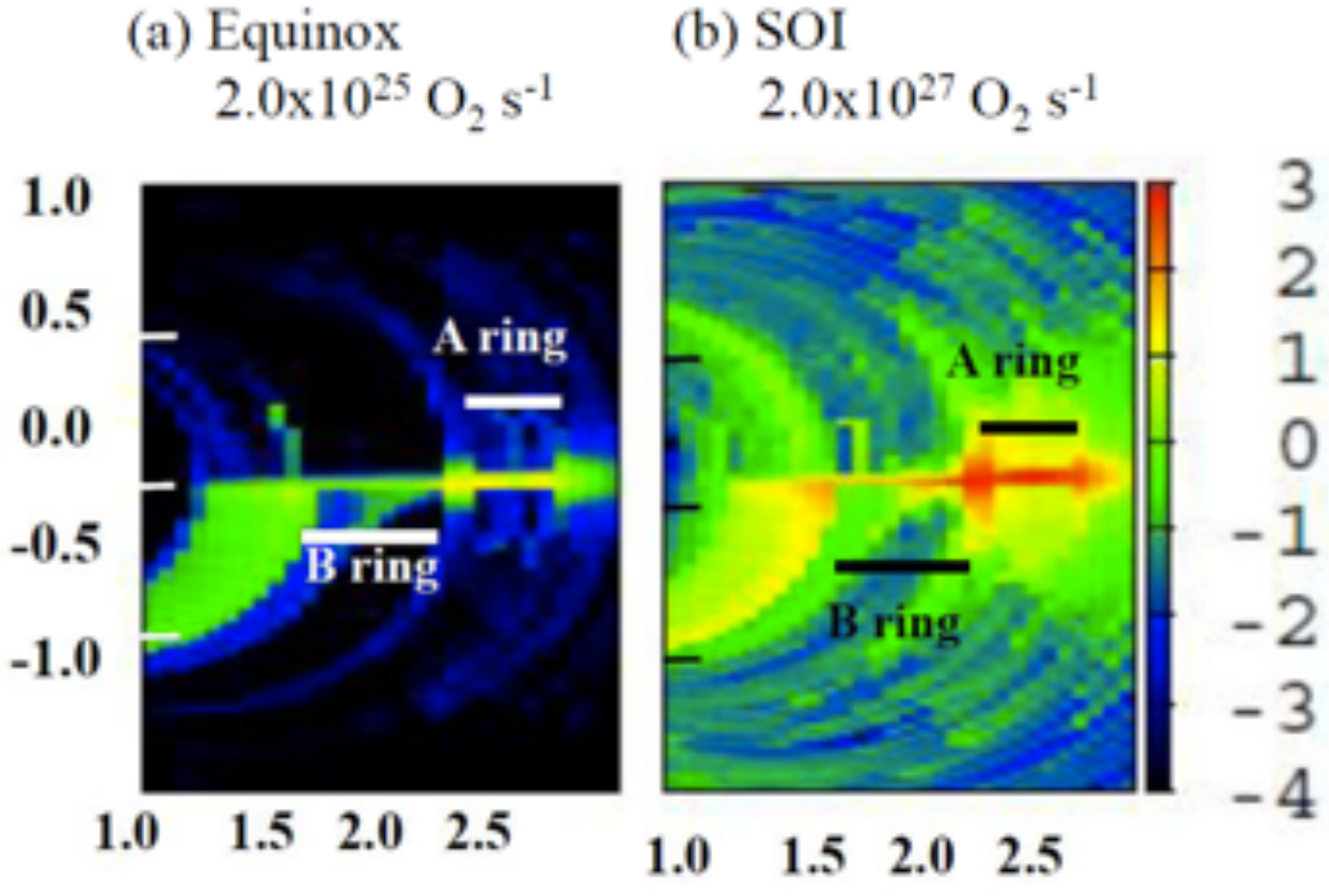

Figure 2: The $O_2^+$ ion density in Log 10 scale as indicated in a color bar. X-axis is the radial distance ($R_S$). Z-axis is the vertical distance ($R_S$). (a) equinox phase: photolytic source rate: 2.0 $x10^{25}$ $s^{-1}$ (b) SOI phase: source rate: 2.0x $10^{27}$ $s^{-1}$.

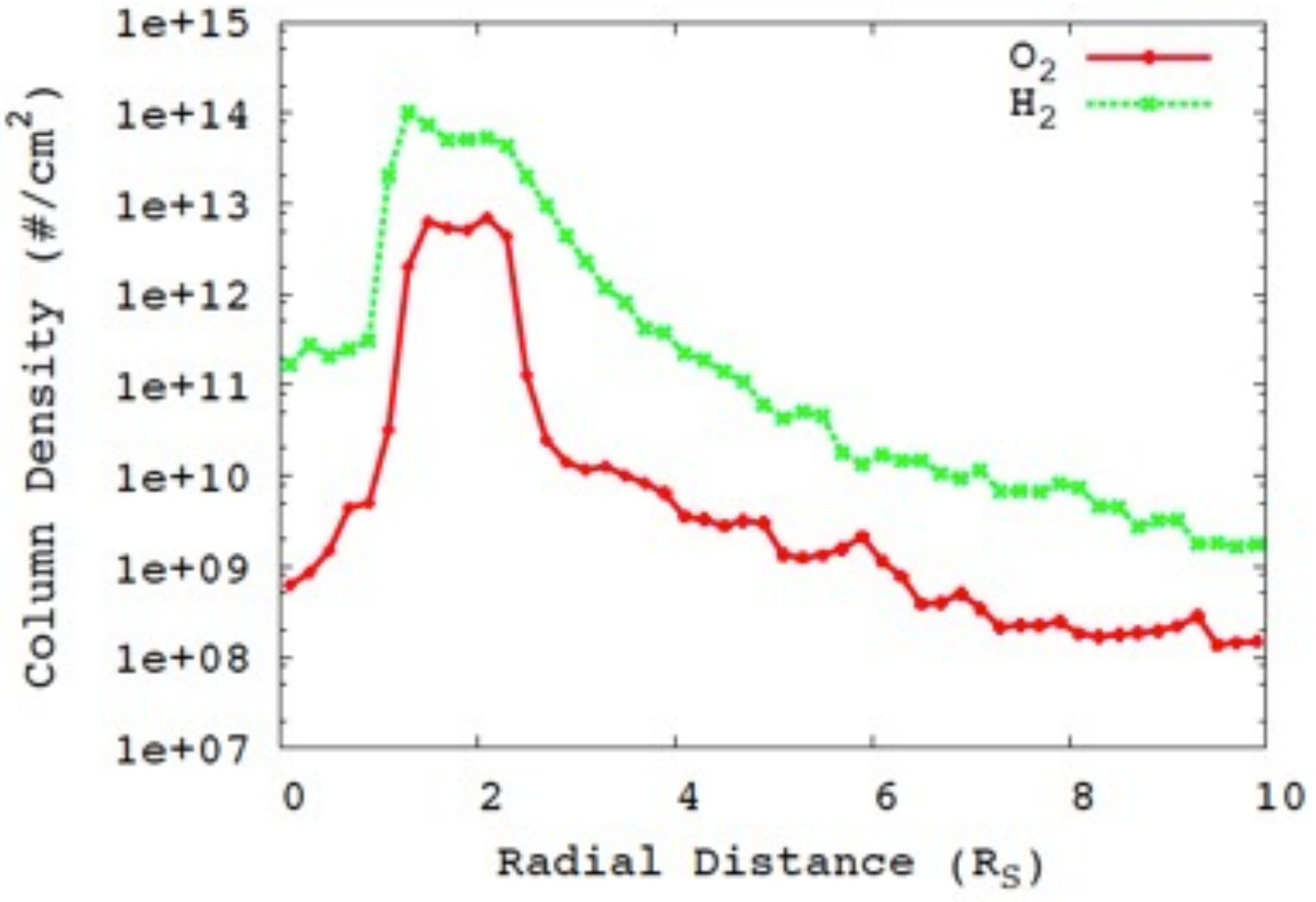

Figure 3: The radial distributions of the neutral $O_2$ and $H_2$ column density with a source rate of ~ 2x10$^{26}$ $H_2$ s$^{-1}$ and of ~1x10$^{26}$ $O_2$ s$^{-1}$, respectively, which ignores the enhancements in Table 1.

**3. Ion Chemistry Model**

As mentioned before, in Elrod et al. (2011) we have shown that there were significant temporal variations in the plasma density and composition in the region 2.5 to 3.5 $R_S$ over the period 2004 to 2010. We also discussed the possible explanations such as the seasonal variability in the ring atmosphere, variability in the Enceladus source rate, and variability in solar activity. To further understand the CAPS observations and examine the roles of Enceladus and of the ring atmosphere source we developed a one-box ion chemistry model to simulate the complex and highly variable plasma environment that was observed by the Cassini CAPS. Some preliminary results for the model described here were reported in Elrod et al. (2011). The plasma chemistry between 2.5 and 3.5 $R_S$ near Saturn's equatorial plane is described by the following parameters: the neutral densities, the diffusion loss timescale, and the density and temperature of hot electrons. The ionization, recombination, ion-molecule reactions and charge exchange for the major neutral species (neutral $H_2O$, OH, O, H, $H_2$ and $O_2$) are also included. The output is the steady-state plasma composition ($H_3O^+$, $H_2O^+$, $OH^+$, $O^+$, $H^+$, $H_2^+$, $O_2^+$ as well as the thermal electron density). The thermal electron density is a sum of the ion densities. The photo-ionization rates are from the Huebner et al., (1992). The cross-sections for electron impact ionization are based on Itikawa et al. (2001; 2009), Kim and Desclaux (2002), Yoon et al., (2008) and Shah et al., (1987). Maxwellian energy distributions are assumed in calculating the electron-ion and electron-neutral reaction rates. In this region, the suprathermal electron density is nearly depleted (Schippers et al., 2008). But there is a small population of the intermediate hot electrons of $T_e$ = 20-30 eV and 40-50 eV which are mainly from photoionization of water-group neutrals (Schippers et al., 2009). This hot-electron population is set to be 0.1 $cm^{-3}$ (Schippers et al., 2009). The thermal electrons, initially assumed to have $T_e$ = 2.0 eV (e.g. Gustafsson and Wahlund, 2010) were shown to primarily affect the electron-ion recombination rates, which are derived from Schreier et al., (1993). Ion-molecule reaction rates are calculated with energy-dependent cross-sections when available with the relative speed determined by both the plasma corotation velocities and the keplerian velocities of neutrals at 3.0 $R_S$. Energy independent reaction rates are used when the data is insufficient. Here, applying a Maxwellian energy distribution for the ions in the ion-molecule reactions is not so important since most of reaction rates are energy independent (as seen in Appendix 1). We also find that using a Maxwellian energy distribution results in only ~20% difference for the energy-dependent cross-sections involving collisions between water-group neutrals and ions.

The rates (listed in Appendix 1) are used in a set of continuity/chemical rate equations for each ion species, $\frac{dn_i}{dt} = S_i - L_i$ with the source rate, $S_i$, and the loss rate, $L_i$. Using the neutral density distributions from the ring atmosphere and the Enceladus torus, the initial values of densities of each ion species were assigned to be zero and the densities at steady state were solved for using various test parameters. The ion sources ($S_i$) are photoionization, electron-impact ionizations and ion-molecule reactions with the neutrals. The loss ($L_i$) to ion-molecule reactions and recombination are also included and the radial diffusion is treated as a constant loss rate. The coupled set of equations is solved iteratively using the 4$^{th}$ order Runge-Kutta method to reach the steady-state.

### 3.1 Modeling at SOI

For the electron properties described above, photoionization is the primary ion source in this region with the hot-electron impact ionization accounting for ~30%. Loss by radial diffusion (~ 2 months; Rymer et al., 2008) was found to be dominated by recombination in agreement with the statement in Sittler et al. (2008): ‘recombination loss for molecular ions dominated in inner magnetosphere inside Dione while diffusion loss dominated in outer magnetosphere'. In our model, recombination is the primary loss process for $H_3O^+$ and $O_2^+$, whereas ion-molecule reactions and recombination compete for the loss of $H_2O^+$ and $OH^+$. Atomic ions, like $O^+$ and $H^+$, are neutralized directly in ion-molecule collisions or indirectly by ion-molecule reactions in which a molecular ion is formed and subsequently lost by recombination.

To understand the situation at SOI, we consider a neutral background of O, OH and $H_2O$ supplied by the Enceladus torus with an average source rate of $1 \times 10^{28}$ $H_2O$ $s^{-1}$ from Cassidy and Johnson (2010) and the $O_2$ and $H_2$ clouds scattered from the ring atmosphere as discussed in Section 2. Since there was no detection of $H_2^+$, we assumed an arbitrary enhancement factor of 5 for the neutral $H_2$ ring source rate due to recycling. This is a reasonable assumption, since at an enhancement factor as high as 20 suggested for $O_2$, CAPS would have detected $H_2^+$ at SOI (Tseng et al., 2011). Whereas the Enceladus torus dominates the total neutral column densities in this region (Elrod et al., 2011), the significant difference in scale heights (ring atmosphere~ 1,000 km for $O_2$ and ~4,000 km for $H_2$; Enceladus torus water products ~6,500km) are such that the near equatorial density is dominated by the ring atmosphere at SOI. The average neutral densities near the equator around 3.0 $R_S$ are shown in Table 2. At SOI, the averaged photoionization rates (Huebner et al., 1992) are applied since the solar activity was average. Using these densities, we find that $O_2^+$ is the dominant ion in this region with a density ~150 $cm^{-3}$ about the same order of magnitude as the average CAPS data ~500 $cm^{-3}$ (Elrod et al., 2011). The total water-group ion density calculated is ~ 20 $cm^{-3}$ as compared to the CAPS data (~150 $cm^{-3}$). $H_3O^+$ is the major water-group ion in our model because of the efficient ion-molecule reactions at the relative speeds considered here. $H_3O^+$ is primarily formed by the reaction of $H_2O^+$ with $H_2$ where the $H_2$ component from the ring source is significant at SOI. Although both densities are lower than the average CAPS data in this region, the ratio of $O_2^+/W^+$ (~7.5) is higher than the ratio found in CAPS SOI data (~2.5-4.5). Therefore, our basic model using average conditions requires modification as discussed below and in Section 3.3. The $O_2^+$ and $W^+$ ion densities and the ratio of $O_2^+/W^+$ from the CAPS SOI data between 2.4 and 2.8 $R_S$ (see Fig. 5 and 6 in Elrod et al., 2011) also appear to increase toward Enceladus. These increases are likely due to spacecraft trajectory as Cassini was passing through the ring plane from the northern side to the southern side.

Overall, the total ion densities at SOI in the above model are about 4 times lower than the CAPS data. This also suggests that the plasma environment at SOI differs from the average conditions described above. The effect of solar activity on the neutral density and, hence, on ion formation are mostly included in our scaling of the source rate. However, we do note that the total ion density obtained from Voyager 2 data near equinox in 1981 at solar maximum was intermediate to the CAPS data at SOI, which occurred at solar average, and the CAPS data near equinox, which occurred at solar minimum (Elrod et al., 2011). This suggests that solar activity might play an important role. Related to this, we note that Saturn's magnetosphere was very compressed at SOI as a result of an 'active solar wind' and the total electron energy at SOI was ~25% higher than the average used above (Rymer, private communication). Hot plasma injection was also identified during SOI, which is indicative of tail reconnection due to significant compression and is related to the corotating interaction (CIR) region (Bunce et al., 2005). Although the data averaged over a number of orbits shown in Schippers et al. (2008) suggested that the suprathermal electron is nearly depleted in inner magnetosphere, the hot electron temperature and fraction can vary significantly in inner magnetosphere (e.g. Young et al., 2005; Moncuquet et al., 2005). Young et al. (2005) found that, at SOI, the hot electron temperature was a few hundred to a few thousand eV, indicating a longitudinal or temporal variation. They also found the hot electron temperature increased with decreasing radial distance and the fraction of hot electron ranged from 0.01% to 5% within 3-5 $R_S$. This component of hot electrons was neglected in the average plasma torus model above. In addition, Fleshman et al. (2010) modeled the physical chemistry of the Enceladus torus (~4 $R_S$) and found that the hot electrons (energy up to 250 eV) do not exceed 1% of the total population. Therefore, it is reasonable to include a component of hot electrons with energy ~400 eV and density ~5.0 $cm^{-3}$ (the fraction ~1%) into the above model, in which case the hot-electron impact ionization will dominate the photoionization. Including this component, the $O_2^+$ ion density becomes ~ 510 $cm^{-3}$ and the $W^+$ ion density becomes ~130 $cm^{-3}$. Both are in remarkably good agreement with the averaged CAPS data ($O_2^+$ ~ 500 $cm^{-3}$ and $W^+$ ~150 $cm^{-3}$). This agreement also appears to confirm that the electron environment in this region was hotter at SOI resulting in the $O_2^+$ density being dominant.

Table 2: The Neutral Background for SOI Case

| Species | Equatorial density ($cm^{-3}$) | Source (reference) |
|---|---|---|
| $H_2O$ | 2000 | Enceladus torus with an average source rate of $1x10^{28}$ $H_2O\ s^{-1}$ (Cassidy and Johnson, 2010) |
| OH | 1000 | Enceladus torus (Cassidy and Johnson, 2010) |
| O | 1000 | Enceladus torus (Cassidy and Johnson, 2010) |
| $H_2$ | 1000 | Photodissociation of Enceladus water torus (Tseng et al., 2011) |
| $O_2$ | 20,000 | Ring atmosphere with an $O_2$ source rate of $2x10^{27}\ s^{-1}$ referred to CAPS data at SOI |
| $H_2$ | 250,000 | Ring atmosphere with a $H_2$ source rate $1x10^{27}\ s^{-1}$ with an enhancement factor of 5 due to recycling (Tseng et al., 2011) |
| Immediate-Hot Electron | 0.1 | Photoelectron from ionization of water-group neutrals (Schippers et al., 2009) |

### 3.2 Modeling at Saturn Equinox

When neutral $O_2$ and $H_2$ source rates from ring atmosphere decrease to of the order of ten percent of the SOI source rate, the neutral $O_2$ and $H_2$ density near the equator drops to roughly one percent of the value at SOI. Therefore, it is surprising that $O_2^+$ is seen in this region by CAPS as Saturn approaches equinox. However, using the model described above $O_2^+$ is also formed in this region of the Enceladus torus by ion-molecule reactions. The primary pathway for formation of $O_2^+$ at equinox is the reaction of $OH^+$ with O.

Using the intermediate source rate in Table 1, we obtain an average $O_2^+$ density ~10 $cm^3$ and a total water-group ion density ~20 $cm^{-3}$. Near equinox, not only does the photolytic $O_2$ source rate drop, but the solar activity was at a minimum corresponding to reduced photoionization rates for solar minimum (Huebner et al., 1992). Reducing the photolytic source rate, the $O_2^+$ density becomes ~3 $cm^3$ and the total $W^+$ density becomes ~15 $cm^{-3}$ in the region of interest near equinox. Results obtained using both an intermediate phase and equinox ring atmosphere are roughly consistent with the average plasma densities observed by CAPS between 2005-2010 (Elrod et al., 2011). In addition to the consistency with the temporal variation in the ion densities, our model shows that the main ion species is $O_2^+$ at SOI while $W^+$ dominates near equinox, which is also in agreement with the CAPS data. The CAPS data also showed that the average ratio, $O_2^+/W^+$ decreased from ~ 2.5-4.5 at SOI to ~<1 near equinox. Although the scatter in the ratios for the later passes is significant this is roughly consistent with our model results.

Because the relative ion-neutral collision speeds are relatively low in this region of the magnetosphere, ion-molecule reactions play an important role on the ion composition favoring for formation of $H_3O^+$ and $O_2^+$ as seen above. The low relative speeds lead to relatively large cross-sections but also significant scattering of the ions. Test simulations without ion-molecule collisions were carried out for comparison and are shown in Table 3. It is clearly seen that the ion-molecule reactions significantly alter not only the ion composition but also the total ion densities. Obviously, the predicted seasonal variation of the ring atmosphere (Tseng et al. 2010) has an impact on the seasonal variations of plasma environment observed by CAPS. However, the $O_2^+$ density is not as sensitive to the variations of neutral $O_2$ density as one might expect since the reaction of $OH^+$ + O in the Enceladus torus can produce a small amount of $O_2^+$. When ion-molecule reactions are included, the total ion densities decrease because the molecular ions, $H_3O^+$ and $O_2^+$, are end-products that are more quickly lost to recombination than atomic ions. In fact, the ion-electron recombination rate of $H_3O^+$ for Te ~ 2 eV is about one order of magnitude higher than other molecular ions accounting for a higher ratio of $O_2^+$ to $W^+$ at SOI. Therefore, the results presented here are very sensitive to $T_e$.

Table 3: The ion densities produced by with/out ion-molecule reactions

| Case | No ion-molecule reactions | With ion-molecule collisions (no hot electron with Te=400 eV) |
|---|---|---|
| SOI | $W^+$ ~ 650 cm $^{-3}$ (mainly $O^+$)<br>$O_2^+$ ~ 10 cm $^{-3}$ | $W^+$ ~ 20 cm $^{-3}$ (mainly $H_3O^+$ and $H_2O^+$)<br>$O_2^+$ ~ 150 cm $^{-3}$ |
| Equinox | $W^+$ ~ 40 cm $^{-3}$ (mainly $O^+$)<br>$O_2^+$ ~ 0.01 cm $^{-3}$ | $W^+$ ~ 15 cm $^{-3}$ (mainly $H_3O^+$ and $H_2O^+$)<br>$O_2^+$ ~ 3 cm $^{-3}$ |

**3.3 $O_2^+/W^+$ ratio at SOI**

As shown above, in absence of the hot electron component with a $T_e$=400 eV, we calculate total ion densities at SOI that are lower than the CAPS data, especially for the $W^+$ ions. Although the inclusion of a hot electron component appears to resolve this, we explore other possibilities for the lower model ion densities and the high $O_2^+$ /$W^+$ ratio at SOI. First, since Smith et al. (2010) found that Enceladus' plumes were likely variable, the contribution of the Enceladus neutral torus could be much higher at SOI not only affecting the O and OH deposited onto the A-ring, but also producing more water-group neutrals in the region of interest. In addition, we ignored the scattered atomic oxygen from dissociation of ring $O_2$ atmosphere. Third, the interaction of returning hydrogen on the ring particles, affecting the ring $H_2$ atmosphere enhancement, is uncertain, which in turn can affect the $W^+$ density through ion-molecule collisions.

To see if the Enceladus' plumes were much more active at SOI, the broad oxygen cloud with a total number of O atoms inside 10 $R_S$ observed by Cassini UVIS before SOI (Esposito et al., 2005; Melin et al., 2009) was used estimate the Enceladus source rate at that time. Assuming that the oxygen atoms are mainly from dissociation of $H_2O$ from the Enceladus' plumes, then ~$2.7x10^{34}$ O atoms would be present in steady state if the Enceladus source rate is ~$0.85x10^{28}$ $s^{-1}$ (Cassidy, in private communication). This estimate is within about 20% of the ~$3.1$-$3.4x10^{34}$ O atoms reported by UVIS, so that the Enceladus plume source at SOI appears to be reasonably close to what is considered to be the average source rate.

Another possible effect on the water-group ion density is the oxygen atoms scattered out of the ring atmosphere, which were not included in Tseng et al. (2010) calculation nor in the correction for the ring particles temperature calculated here. Based on our model of the ring $O_2$ atmosphere, the oxygen atoms are mainly produced by photodissociation of $O_2$ and charge transfer between $O^+ + O_2$. Due to rapid loss to the ring absorption, which also contributes to the $O_2^+$ enhancement discussed earlier, the ring component of the atomic oxygen density is ~ 2,000-4,000 $cm^{-3}$ near equator at 2.5-3 $R_S$. Although this is slightly larger than the oxygen population in the Enceladus' water torus (~1,000 $cm^{-3}$ in Cassidy and Johnson, 2010), the additional O increases the $W^+$ ion density by only ~10% when photoionization dominates.

As mentioned before, the magnetosphere at SOI was very compressed (e.g. Bunce et al., 2005) so that the SOI plasma environment differs from the average environment. Therefore, including a component of hot electrons with a temperature Te=400 eV and density Ne=5 $cm^{-3}$ (~1% of the total electron density), resulted in ion densities and an $O_2^+/W^+$ ratio that were in good agreement with the SOI data as discussed in Sec. 3.1. The hot electrons affect the ratio by producing more $W^+$ ions (compared to $O_2^+$) than photoionization (see reaction rates in Appendix 1). Interestingly, with a hot electron component, the $O_2^+/W^+$ ratio becomes insensitive to the very uncertain $H_2$ enhancement factor. Vice versa, for a fixed $H_2$ density, the total ion density decreases and the $O_2^+/W^+$ ratio increases when the hot electron component decreases.

**4. Plasma Temperature**

Our analysis of the CAPS data showed that there were not only the drastic variations in the ion densities and composition from 2004 (SOI) to 2010, but also a significant drop in the ion temperature from SOI to the 2005-2010 era (Elrod et al., 2011). At SOI the ion temperatures were close to pickup energy, but were, roughly, less than half that for 2005-2010. Although there were considerable uncertainties in the plasma flow speeds obtained from our 1D fit to the data, which could affect the temperature estimates, the flow speeds were reasonably close to, or slightly larger than, the corotational velocity for all the examined passes. Following the pick-up of newly formed ions, the ion temperature can be modified by a number of processes including diffusion and the interactions with the electrons and neutrals. Therefore, to understand the low ion temperatures detected by CAPS, we simultaneously solved the set of continuity equations for density *and* energy (temperature) for each ion species and the thermal electron. That is, we added the energy equations into the ion-chemistry model in Section 3.

Once ionized by the photons, electrons and charge exchange reactions, an ion newly formed near the equator is picked-up: i.e., it is accelerated to the corotation speed, $V_{co}$ , and attains a gryomotion with an energy $E_P = \frac{1}{2} m_i Vrel^2$ where $Vrel$ (= $Vco - Vorb$) with $V_{orb}$ the local Keplerian speed. Since the distribution isotrophizes fairly rapidly by wave particle scattering and mutual collisions, the motion about the mean flow is often characterized by a temperature, $kT_i \sim E_P$. Using this as the initial thermal energy, the energy loss processes are considered. One such process is the Coulomb interactions between ions and electrons. The suprathermal electron population (a few hundred to a few thousand eV) in this region is, typically, negligible but a small population of intermediate-hot electrons, the fresh photoelectrons, is usually present (Schippers et al., 2009). However, as mentioned before, a population of hot electron with Te~100-3,000 eV was detected at SOI with a fraction that varied between 0.01% and 5% within 3-5 $R_S$ (e.g., Young et al., 2005). The thermal equilibration rate due to Coulomb interactions between ions and electrons is given by

$\gamma_\beta^\alpha = 1.8 \times 10^{-19} \frac{(m_\alpha m_\beta)^{\frac{1}{2}} Z_\alpha{}^2 Z_\beta{}^2 n_\beta \lambda_{\alpha\beta}}{(m_\alpha T_\beta + m_\beta T_\alpha)^{3/2}}$ where $\lambda_{\alpha\beta}$ ~10-20 is the Coulomb logarithm [Book, 1990]. Energy loss from the ion population also occurs by ion-electron recombination, by radial diffusion, and by momentum transfer during ion-molecule collisions. These aspects are all accounted for in the following.

To approximate the momentum transfer in ion-molecule reactions, the *reaction rates* between ions and neutrals are determined by the Langevin rate constant, $2\pi[\alpha(Ze)^2/m_i]^{0.5}$ where $\alpha$ is the polarizability of neutrals (e.g. Johnson et al., 2006). This rate constant, applicable for slow collisions, is a measure of the size of the region for which the ion and neutral interact strongly and roughly share the center of mass energy. For near-equal mass collisions this implies that about half the ion energy lost to neutrals (see Appendix 2).

The thermal electron energy is determined by cooling to the neutral water molecules, energy loss to radial transport and recombination, and heating by the Coulomb interactions with ions and hot electrons. A new electron formed by photoionization is given an excess energy obtained from Huebner et al., (1992). Due to very large electron-impact cross sections for rotational and vibrational excitation of polar molecules, cooling only by water molecules is taken into account. This is consistent with what occurs in cometary ionospheres (Cravens and Korosmezey, 1986). Although neutral $O_2$ is dominant at SOI, electron cooling to $O_2$ is negligible, since the energy loss cross sections are about two orders-of-magnitudes smaller than those for $H_2O$ at Te ~ 2 eV. The cooling rates with $H_2O$ are assumed to be half the values used in Cravens and Korosmezey (1986) based on a recent review of rotational cross sections (Itikawa, 2007; Cravens et al. 2011).

### 4.1 Modeling Results: Thermal Electron Temperature

We simultaneously solve for plasma densities and energies at both SOI and Equinox using the neutral densities discussed in Section 3. The ion densities attained are about the same as the values given in Table 3 in which the ion temperature was ignored. Our modeling of the thermal electron temperature gives ~1-2.5 eV, which is in good agreement with the Cassini observations (e.g. Tokar et al., 2006; 2008). We also note that the thermal equilibration time between heavy ions (e.g. $W^+$ and $O_2^+$) and electrons is approximately $10^6$ - $10^7$ sec, which is much longer than the plasma lifetime, ~a few x$10^5$ sec, due to recombination. Therefore, Coulomb collision with the heavy ions is not an efficient electron heating mechanism. For heating by the protons the equilibrium time is faster, ~a few x$10^5$ sec, but the proton density is low (~10% of $W^+$ density) so the net effect is again small. The heating of thermal electrons by the co-rotating proton population has been discussed in Rymer et al. (2007) and Sittler et al. (2006). On the other hand, we find that the equilibration time between thermal electrons and the hot electrons is ~ $10^3$ - $10^4$ sec, which makes this an important heat source, with electron-water cooling the dominant loss process. Gustafsson and Wahlund (2010) and Cravens et al. (2011) had the similar conclusions in their studies of the electron energy in Saturn's inner magnetosphere and in Enceladus' water torus.

### 4.2 Modeling Results: Ion Temperature

Using the above model which includes the hot electron component, we obtain thermal energies for $O_2^+$ of ~32 eV at SOI and ~12 eV at equinox as well as $W^+$ energies of ~20 eV at SOI and ~ 11 eV at equinox. These average energies are close to ~80-100% of normal pickup temperature (~20 eV for $W^+$ ions and ~40 eV for $O_2^+$ ions at 3.0 $R_S$) for SOI conditions and ~30-70% of pickup temperature energy for both intermediate and equinox ring atmosphere conditions. Since cooling to the electrons is inefficient, as discussed above, the low ion temperatures are mostly a result of the energy exchange during ion-molecule collisions. These compare favorably with the CAPS data which indicated that the ions were near pickup energy at SOI but near equinox were ~50% of pickup energy (Elrod et al., 2011).

From this analysis it appears that the momentum transfer during low relative-speed ion-molecule collisions might account for the low ion temperatures detected by CAPS. Johnson et al., (2006) also calculated the scattered ion energy distribution in the corotating frame for very low relative-speed collisions of $O_2^+$ ions and neutral $O_2$ molecules over the main rings. The result, shown in their Fig. 2, deviated significantly from an isotropic Maxwellian energy distribution so more detailed simulations are needed in future work. For example, the significant ion temperature anisotropy has been ignored but is clearly important requiring a 3D analysis of the CAPS data that is now in progress. In addition, calculating the energy exchange between ions and electrons, we assumed a Maxwellian distribution while the fresh pickup ions have a "pancake (or ring)" distribution (e.g. Tokar et al., 2008). This may in part explain the modeled electron temperature being slightly lower than the detected values as suggested by Gustafsson and Wahlund (2010) and Cravens et al. (2011).

### 4.3 Interactions with Dust

The region of interest, between 2.5 and 3.5 $R_S$, is permeated with dust particles from the F, G and E rings. The F ring, a narrow structure just outside of the A ring, is dynamic, changing rapidly both spatially and temporally, even on time scales of hours (Murray et al., 2008). One of Saturn's mysterious rings, the G ring, is ~2.8 $R_S$ from the center of planet, the outer edge or our region. The entire G ring could be derived from an arc of debris held in resonance with Mimas and produced by micrometeorite impacts generating ejecta that subsequently spreads out (Hedman et al., 2007). The depletions of energetic electrons detected by Cassini MIMI suggested that a lot more small grains are distributed within the ring than is seen by Cassini cameras (Roussos et al., 2005). The diffusive E ring is populated by the icy particles and dust emanating from Enceladus‘south pole. All these rings are composed of sub-micron and micron-sized grains (e.g. Hedman et al., 2008; Kempf et al., 2008; Horanyi et al., 2009). The Cassini CDA data showed the dust density of grains larger than 0.8 micron is ~ $2x10^{-3}$ $m^{-3}$ near Mimas‘ orbit and the peak density at ~4.0 $R_S$ is ~0.2 $m^{-3}$ (Kempf et al., 2008). However, the dust density of smaller particles would be much higher since the dust population follows a steep size distribution with $r_d^{-a}$ where $r_d$ is the grain radius ($a$~4-5 from Kempf et al., 2008).

In this dusty plasma, the grains are charged a few volts negatively (Jurac et al., 1995; Kempf et al., 2006). Charged grains larger than micron-sized in the E ring have velocities comparable to the local Keplerian velocity (e.g. Hamilton and Burns, 1994). The smaller charged grains have velocities between the local Keplerian velocity and the corotational plasma velocity. Because the Debye length is generally around a few meters, which is smaller than the inter-grain distance, the grains act as isolated screened particles. The momentum transfer between the ions and the charged dust occurs at a rate (~ $\sigma n_d V_{rel}$). *$\sigma$ is the screened Coulomb cross-section, which at these electron densities is ~ $2x10^{-6}$ $cm^2$.* With an average dust density ($n_d$) ~ $2x10^{-9}$ $cm^{-3}$ and relative velocity, $V_{rel}$ ~$1.5x10^6$ cm $s^{-1}$ at 3.0 $R_S$, the ion-dust interaction time is ~$2x10^8$ s. Since this is much longer than the ion lifetime, ~ a few times $10^5$ s due to recombination, ions do not even experience one interaction with the dust prior to their loss. Therefore, ion-dust collisions can not account for the low ion temperatures detected in CAPS data in 2005-2010.

## 5. Summary

To understand the surprisingly large changes in the ion density, temperature and composition between solstice and equinox revealed by Cassini CAPS (Elrod et al., 2011), we carried out simulations of seasonal variations in Saturn's plasma environment near the equatorial plane between 2.5-3.5 $R_S$. We used a revised ring atmosphere model in which we accounted for the temporal change in the average ring particle temperature as well as the change in the solar incident angle with respect to the ring plane. This resulted in a more drastic seasonal variation than our previous prediction in Tseng et al. (2010) although there are still considerable uncertainties on the importance of oxygen recycling on the surfaces of the ring particles. Combining the water-group neutrals from Enceladus torus and the $O_2$ and $H_2$ molecules from the ring atmosphere, our one-box ion-chemistry model can reasonably account for the large variations in the ion densities and composition between SOI and Equinox observed by CAPS if we assume the magnetosphere was highly compressed so that a hot component of electrons was present at SOI. It also suggests that the relatively low-speed ion-molecule reactions in this region produce measurable amounts of $O_2^+$ even with a severely quenched ring atmosphere near equinox. This does not take place in the magnetosphere outside Enceladus because of increasing the relative collision speed. However, there are still other constraints on this model such as interactions with dust grains, neutral sources from radiolytic decomposition of small icy dust in this region, and so on. These will be explored in more detail in future. Also, the modeling described here confirms our conclusion in Elrod et al. (2011) that the observed changes in the plasma within that time period are primarily due to predicted seasonal variations (Tseng et al. 2010) in the ring atmosphere. We showed that:

1. Including the orientation of the ring plane to the Sun and the seasonal variation of ring particle temperature, the neutral $O_2$ source rate for the ring atmosphere is ~$2.0 \times 10^{27}$ $s^{-1}$ at SOI (solstice) by scaling to CAPS data (Elrod et al., 2011). Near equinox, it is ~~$2.0 \times 10^{25}$ $O_2$ $s^{-1}$ allowing for the same enhancement factor due to recycling.
2. $O_2$ and $H_2$ scattered from the ring atmosphere by ion-molecule collisions are important sources for the magnetosphere (Johnson et al., 2006; Martens et al., 2008; Tseng et al., 2010) as are the water products from Enceladus neutral torus (Cassidy and Johnson 2010). These sources are included in an ion-chemistry model to describe the temporal changes in ion densities, temperatures and composition detected by CAPS. Our results show that, at SOI, the $O_2^+$ ion density is ~ 510 $cm^{-3}$ and the $W^+$ ion density is ~130 $cm^{-3}$. We also find that $O_2^+$ density is ~3 $cm^3$ and a total $W^+$ density is ~15 $cm^{-3}$ near equinox. Both are in good agreement with the averaged CAPS data. Therefore, in the region between 2.5-3.5 $R_S$, although the possible variability in the Enceladus source might contribute, the observed variations in plasma densities and composition were primarily seasonal as predicted.
3. The ion-neutral collisions, due to low relative speed in this region, play important roles on the observed composition and the surprisingly low ion temperatures found in the plasma data close to equinox. Based on the dust density in this region (Kempf et al., 2008), the ion-dust interaction appear to be negligible.
4. Our model indicates that the presence of hot electrons at SOI (Te=400 eV) plays a dominant role in determining both the ion densities and composition. In the absence of a significant hot electron component photoionization dominates and the $H_2$ from the ring atmosphere becomes important in determining the $W^+$ density. In spite of the larger scale height over the ring plane for the $H_2$ than that for $O_2$, CAPS should have detected $H_2^+$ ions in the Cassini division at SOI if $H_2$ and $O_2$ are formed in the ring atmosphere with a stoichiometric ratio of 2:1 and the enhancements due to recycling were similar (Tseng et al., 2011). Therefore, the actual $H_2$ density in our region remains uncertain requiring continued analysis of CAPS data.

In spite of the uncertainties mentioned, it is exciting that the predicted seasonal variations in the ring atmosphere appear to account for the significant differences in Saturn's trapped plasma observed by the CAPS instrument on Cassini and earlier by the plasma instrument on Voyager.

**Appendix 1: Reaction Rates**

| | Reaction | Rate constant ($s^{-1}$) (Huebner et al., 1992)<br>Above: Solar Min<br>Below Solar Max |
|---|---|---|
| Photoionization | $H_2O + h\nu \rightarrow H_2O^+ + e$ | 3.55D-9<br>8.9D-9 |
| | $H_2O + h\nu \rightarrow OH^+ + H + e$ | 5.95D-10<br>1.62D-9 |
| | $H_2O + h\nu \rightarrow O^+ + H_2 + e$ | 6.28D-11<br>2.37D-10 |
| | $H_2O + h\nu \rightarrow H^+ + OH + e$ | 1.41D-10<br>4.37D-10 |
| | $OH + h\nu \rightarrow OH^+ + e$ | 2.65D-9<br>7.0D-9 |
| | $O + h\nu \rightarrow O^+ + e$ | 2.15D-9<br>5.8D-9 |
| | $H + h\nu \rightarrow H^+ + e$ | 4.8D-9<br>1.85D-9 |
| | $O_2 + h\nu \rightarrow O_2^+ + e$ | 4.3D-9<br>1.27D-8 |
| | $O_2 + h\nu \rightarrow O^+ + O + e$ | 1.3D-9<br>3.73D-9 |
| | $H_2 + h\nu \rightarrow H_2^+ + e$ | 9.1D-10<br>1.23D-9 |
| | $H_2 + h\nu \rightarrow H^+ + H + e$ | 4.0D-10<br>3.0D-10 |

| Electron-impact ionization | Reaction | Rate constant $cm^3\ s^{-1}$ (references)<br>Top: Te=2.0 eV<br>Middle: Te=25.0eV<br>Bottom: 400.0 eV |
|---|---|---|
| | $O_2$ + e -> $O_2^+$ +2e | 8.4D-12 (Itikawa, 2009)<br>3.1D-8<br>1.2D-7 |
| | $O_2$ + e -> $O^+$ + O +2e | 1.2D-13 (Itikawa , 2009)<br>1.2D-8<br>6.5D-8 |
| | $H_2O$ + e -> $H_2O^+$ +2e | 1.4D-11 (Itikawa and Mason, 2005)<br>3.0D-8<br>8.8D-8 |
| | $H_2O$ + e -> $OH^+$ +H +2e | 3.0D-13 (Itikawa and Mason, 2005)<br>7.9D-9<br>2.9D-8 |
| | $H_2O$ + e -> $O^+$ +$H_2$ +2e | 2.4D-15 (Itikawa and Mason, 2005)<br>8.5D-10<br>4.6D-9 |
| | $H_2O$ + e -> $H^+$ +$O_2$ +2e | 5.6D-14 (Itikawa and Mason, 2005)<br>5.0D-9<br>2.5D-8 |
| | OH + e -> $OH^+$ +H +2e | 1.4D-11 [a]<br>3.0D-8<br>8.8D-8 |
| | OH + e -> $O^+$ +H +2e | 3.0D-13 [a]<br>7.9D-9<br>2.9D-8 |
| | OH + e -> $H^+$ +O +2e | 5.6D-14 [a]<br>5D-9<br>2.5D-8 |
| | O + e -> $O^+$ +2e | 6.7D-12 (Kim and Desclaux, 2002)<br>3.0D-8<br>1.0D-7 |
| | H + e -> $H^+$ +2e | 1.2D-12 (Shah et al., 1987)<br>1.5D-8<br>2.7D-8 |
| | $H_2$ + e -> $H_2^+$ +2e | 4.7D-12 (Yoon et al., 2008)<br>2.1D-8<br>4.5D-8 |
| | $H_2$ + e -> $H^+$ +H +2e | 1.8D-15 (Yoon et al., 2008)<br>1.2D-9<br>3.3D-9 |
| Charge transfer collisions | Reaction | Rate constant $cm^3$ $s^{-1}$ (references) |
| | $H_2O$+ + $H_2O$ -> OH + $H_3O^+$ | 7.5D-10 (Lishawa et al., 1990) |
| | $H_2O$+ + $H_2O$ -> $H_2O$ + $H_2O^+$ | 1.5D-9 (Lishawa et al., 1990) |
| | $H_2O^+$ + OH -> O + $H_3O^+$ | 6.9D-10 (Ip, 1997) |
| | $H_2O^+$ + O -> $H_2$ + $O_2^+$ | 4.0D-11 (Ip, 1997) |
| | $H_2O^+$ + $O_2$ -> $H_2O$ + $O_2^+$ | 5.0D-9 (Fehsenfeld et al., 1967) |
| | $H_2O^+$ + $H_2$ -> H + $H_3O^+$ | 8.3D-10 (Ip, 1997) |
| | $OH^+$ + $H_2O$ -> OH + $H_2O^+$ | 1.5D-9 (Lishawa et al., 1990) |
| | $OH^+$ + $H_2O$ -> O + $H_3O^+$ | 1.3D-9 (Ip, 1997) |
| | $OH^+$ + OH -> O + $H_2O^+$ | 7.0D-10 (Ip, 1997) |
| | $OH^+$ + O -> H + $O_2^+$ | 7.1D-10 (Ip, 1997) |
| | $OH^+$ + $O_2$ -> OH + $O_2^+$ | 5.9D-10 (Ip, 1997) |
| | $OH^+$ + $H_2$ -> H + $H_2O^+$ | 1.0D-9 (Ip, 1997) |
| | $O^+$ + $H_2O$ -> O + $H_2O^+$ | 6.1D-9 (Dressler et al., 2006) |
| | $O^+$ + OH -> $O_2$ + $H^+$ | 1.2D-10 (Giguere and Huebner, 1978) |
| | $O^+$ + OH -> O + $OH^+$ | 3.0D-10 (Giguere and Huebner, 1978) |
| | $O^+$ + O -> O + $O^+$ | 1.8D-8 (Ip, 1997) |
| | $O^+$ + H -> O + $H^+$ | 2.0D-9 (Tawara, 1985) |
| | $O^+$ + $O_2$ -> O + $O_2^+$ | 2.0D-9 (Stebbings, 1963) |
| | $O^+$ + $H_2$ -> H + $OH^+$ | 17D-9 (Ip, 1997) |
| | $H^+$ + $H_2O$ -> H + $H_2O^+$ | 7.4D-10 (Lindsay et al., 1997) |
| | $H^+$ + OH -> H + $OH^+$ | 7.4D-10 (Lindsay et al., 1997) |
| | $H^+$ + O -> H + $O^+$ | 7.0D-10 (Ip, 1997) |
| | $H^+$ + H -> H + $H^+$ | 5.9D-9 (Tawara, 1985) |
| | $H^+$ + $O_2$ -> H + $O_2^+$ | 2.1D-9 (Rees, 1989) |
| | $H^+$ + $H_2$ -> H + $H_2^+$ | 3.4D-11 (Tawara, 1985) |
| | $O_2^+$ + $O_2$ -> $O_2$ + $O_2^+$ | 7.4D-10 (Tseng et al., 2010) |
| | $H_2^+ + H_2O$ -> $H + H_3O^+$ | 3.4D-9 (Ip, 1997) |
| | $H_2^+ + H_2O$ -> $H_2 + H_2O^+$ | 3.9D-9 (Ip, 1997) |
| | $H_2^+ + OH$ -> $H + H_2O^+$ | 7.6D-10 (Ip, 1997) |
| | $H_2^+ + OH$ -> $H_2 + OH^+$ | 7.6D-10 (Ip, 1997) |
| | $H_2^+ + O$ -> $H + OH^+$ | 1.0D-9 (Ip,1997) |
| | $H_2^+ + H$ -> $H_2 + H^+$ | 6.4D-10 (Ip,1997) |
| | $H_2^+ + O_2$ -> $H_2 + O_2^+$ | 1.9D-10 (Tseng et al., 2011) |
| | $H_2^+ + H_2$ -> $H_2 + H_2^+$ | 2.6D-10 (Tseng et al., 2011) |
| Recombination | Reaction | Rate $cm^3\ s^{-1}$ (Te=2 eV; Schreier et al., 1993) |
| | $H_3O^+ + e$ -> $H_2O + H$<br>-> $OH + H_2$<br>-> $OH + H + H$ | 3.5D-8<br>2.34D-8<br>6.5D-7 |
| | $H_2O^+ + e$ -> $OH + H$<br>-> $O + H_2$ | 4.5D-8 |
| | $OH^+ + e$ -> $O + H$ | 8.5D-9 |
| | $O^+ + e$ -> $O$ | 4.0D-12 |
| | $O_2^+ + e$ -> $O + O$ | 2.5D-8 |
| | $H^+ + e$ -> $H$ | 1.3D-13 |
| | $H_2^+ + e$ -> $H_2$ | 4.0D-8 |
| Diffusion time | | ~ 2 months (~5.2D6 sec; Rymer et al., 2008) |

[a] For simplicity, we apply the same electron-impact cross-section data of H2O molecules to OH molecules.

**Appendix 2: Energy Change of Pickup Ions After Scattered by Neutrals**

Here we examine the effective temperature, or width of the speed distribution, of an ion that is scattered by a neutral and then re-accelerated to co-rotation speed with a corresponding change in the gyromotion. If the scattered ion is moving with an assumed isotropically Maxwellian energy distribution, the average energy in each direction is $\langle m\mathbf{v_i}^2/2\rangle = kT_i/2$. Below we define the width of the energy distribution in each direction and then get a temperature.

For an ion-neutral collision (with equal mass): For a fast collision in which the interaction is weak and the deflections small: $A^+(v_i) + A(v_o) \rightarrow A(v_i') + A^+(v_o)$. If before the collision, we write $\mathbf{v_i}$

= $\mathbf{v_{co}}$, the ion velocity after picked up is $\mathbf{v_i'} = \mathbf{v_{co}} + \mathbf{u_p}$ abs($\mathbf{v_o}$-$\mathbf{v_{co}}$) with $\mathbf{u_p}$, a randomly oriented unit vector in 2D perpendicular to the local field line assumed to be in z direction.

For a slow collision, for which the Langevin cross sections applies and the interaction is strong, $A^+(v_i) + A(v_o) \rightarrow A((\mathbf{v_i}+\mathbf{v_o})/2 + \mathbf{u_1}\,\mathrm{abs}(\mathbf{v_i}-\mathbf{v_o})/2) + A^+((\mathbf{v_i}+\mathbf{v_o})/2 - \mathbf{u_1}\,\mathrm{abs}(\mathbf{v_i}-\mathbf{v_o})/2)$.

Here $\mathbf{u_1}$ is random in 3D due to the assumed isotropic scattering in the center of mass. We can write $\mathbf{u_1} = \sin\theta\cos\varphi\,\mathbf{x} + \sin\theta\sin\varphi\,\mathbf{y} + \cos\theta\,\mathbf{z}$ with $\mathbf{du_1} = (d\cos\theta\, d\varphi)/4\pi$ where $\theta$ is the angle with the z-axis and $\varphi$ is the azimuthal angle about the z-axis. Therefore, for initial $\mathbf{v_i} = \mathbf{v_{co}}$, the ion velocity after scattering *and* pick-up is $\mathbf{v_i'} = \mathbf{v_{co}} - \mathbf{u_p}\,\mathrm{abs}((\mathbf{v_{co}}-\mathbf{v_o})/2 + \mathbf{u_{1p}}\,\mathrm{abs}(\mathbf{v_{co}}-\mathbf{v_o})/2) + \mathbf{u_{1z}}\,\mathrm{abs}(\mathbf{v_{co}}-\mathbf{v_o})/2$.

After the collisions the width of the energy distribution is $<(\mathbf{v_i'} - \mathbf{v_{co}})^2>$ where $< >$ means average over all the random orientations. Again, assuming that before the collision $\mathbf{v_i} = \mathbf{v_{co}}$ with corotation in the $\mathbf{x}$ direction for a fast collision with weak interaction (like an ionization), then the integral over the random directions is $\int (\mathbf{v_i'}-\mathbf{v_{co}})^2\, \mathbf{du_p} = (\mathbf{v_{co}}-\mathbf{v_o})^2$, or the width in each direction ($\mathbf{x}$ and $\mathbf{y}$) is $(\mathbf{v_{co}}-\mathbf{v_o})^2/2$ so that $kT_i/2 = (m/2)(\mathbf{v_{co}}-\mathbf{v_o})^2/2$ or $kT_i = m(\mathbf{v_{co}}-\mathbf{v_o})^2/2$. On the other hand, for a slow collision with an strong interaction, then

$$\int\int(\mathbf{v_i'}-\mathbf{v_{co}})^2\,\mathbf{du_1}\,\mathbf{du_p} = \int\int[\mathbf{u_p}\,\mathrm{abs}((\mathbf{v_{co}}-\mathbf{v_o})/2 + \mathbf{u_{1p}}\,\mathrm{abs}(\mathbf{v_{co}}-\mathbf{v_o})/2) + \mathbf{u_{1z}}\,\mathrm{abs}(\mathbf{v_{co}}-\mathbf{v_o})/2]^2\,\mathbf{du_1}\,\mathbf{du_p}$$

$$= \int\int[(\mathrm{abs}((\mathbf{v_{co}}-\mathbf{v_o})/2 + \mathbf{u_{1p}}\,\mathrm{abs}(\mathbf{v_{co}}-\mathbf{v_o})/2))^2 + (\mathrm{abs}(\mathbf{v_{co}}-\mathbf{v_o})/2)^2]\,\mathbf{du_1}\,\mathbf{du_p}$$

$$= ((\mathbf{v_{co}}-\mathbf{v_o})/2)^2 \int((1 + 2(\mathbf{u_{1p}})_x + \mathbf{u_{1p}}^2) + 1)\,\mathbf{du_1}\,\mathbf{du_p} = (\mathbf{v_{co}}-\mathbf{v_o})^2/2$$

That is the width is half that above which is the simple result used in our ion energy analysis.

It would be clearer to break down into $\mathbf{x},\mathbf{y},\mathbf{z}$ directions:

$$kT_z/2 = (m/2)\int\int[\cos\theta\,\mathrm{abs}(\mathbf{v_{co}}-\mathbf{v_o})/2]^2\,\mathbf{du_1} = [(\mathbf{v_{co}}-\mathbf{v_o})/2]^2/3 = (1/12)\,(m/2)\,(\mathbf{v_{co}}-\mathbf{v_o})^2$$

$$kT_x/2 = (m/2)\,((\mathbf{v_{co}}-\mathbf{v_o})/2)^2 \int\int \cos\alpha^2\,[(1+\sin\theta\cos\varphi)^2 + (\sin\theta\sin\varphi)^2]\,\mathbf{du_1}\,\mathbf{du_p}$$

$$= (m/2)\,((\mathbf{v_{co}}-\mathbf{v_o})/2)^2 \int(1/2)\,[(1+\sin\theta\cos\varphi)^2 + (\sin\theta\sin\varphi)^2]\,\mathbf{du_1}$$

$$(5/24)\,(m/2)\,(\mathbf{v_{co}}-\mathbf{v_o})^2.$$

Since $kT_x/2 = kT_y/2$, then $3kT_i/2 = [(5/24)+(5/24)+(2/24)]\,(m/2)\,(\mathbf{v_{co}}-\mathbf{v_o})^2 = (m/2)\,(\mathbf{v_{co}}-\mathbf{v_o})^2/2$ as above. In Johnson (2005) and Johnson et al. (2006) cases in which the ion initially has gyromotion are considered. The resulting distribution is seen to be much narrower than a Maxwellian.

**Acknowledgment**

We thank Prof. Wing Ip for helpful suggestions on the plasma interactions. We also thank Dr. T. A. Cassidy in and Dr. Abi Rymer for useful discussions on Enceladus neutral clouds and Saturnian magnetosphere at SOI. This work is supported by a grant from NASA's Planetary Atmosphere's Program and a subgrant from the Cassini CAPS instrument team at SwRI via a grant through JPL.